\documentclass[%
 reprint,
 amsmath,amssymb,
 aps,
pra,
]{revtex4-1}

\usepackage{graphicx}
\usepackage{dcolumn}
\usepackage{bm}

\def\Bra#1{\left<#1\right|}
\def\Ket#1{\left|#1\right>}
\begin{document}
\preprint{APS/123-QED}

\title{Role of different type of sub-systems in doubly driven $\Lambda$-system in $^{87}$Rb}
\author{Kanhaiya Pandey}
 \email{kanhaiyapandey@gmail.com}
\affiliation{%
 Centre for Quantum Technologies, National University of Singapore, 3 Science Drive 2, Singapore 117543.
}%




\date{\today}

\begin{abstract}
The well known $\Lambda$-system using two ground state hyperfine levels, F=1 and F=2 of 5$S_{1/2}$ and one hyperfine level, F=2 of excited state of 5$P_{3/2}$ of $^{87}$Rb has been recently studied using two counter-propagating control lasers \cite{CPN12}. The experiment shows conversion of electromagnetically induced transparency (EIT) into electromagnetically induced absorption (EIA) because the doubly driven $\Lambda$-system forms various sub-systems. We here present detailed theoretical study of the different possible sub-systems created by this configuration. We also explore the possibility of tuning the strength of individual sub-systems by changing the polarization of the control lasers.
\begin{description}
\item[PACS 32.80.Qk, 42.50.Gy, 42.50.Hz]
\end{description}
\end{abstract}
\pacs{Valid PACS appear here}
\maketitle


\section{Introduction}
Laser induced coherence between levels in the multi-level systems is the core of all the quantum
interference effects in near resonant laser-atom interaction. This laser induced coherence is also known as transfer of coherence (TOC), since simultaneous driving of different levels with lasers, induces coherence between the levels which are not directly driven. The TOC gives rise to interesting phenomenons like Electromagnetically Induced Transparency (EIT) \cite{HFI90}, Electromagnetically Induced Absorption (EIA) and Coherent Population Trapping (CPT).
EIT is an example of suppressing the absorption of a probe laser in the presence of
a control laser in the three-level systems ($\Lambda$, $V$  and $\Xi$) due to TOC between levels which are not allowed by dipole transition. EIT has been extensively studied due to its potential application in wide variety of fields such as lasing without inversion \cite{HFI90,GSA91}, high resolution spectroscopy \cite{KPW05,JLM95},
enhancement of second and third order nonlinear processes \cite{IMH89},
polarization control \cite{KAV08,WIG98}, and storage of light \cite{PFM01}.

The modification of the probe laser absorption due to TOC has been investigated beyond three-level system \cite{GWR04,ZKA09,WBO09,BMW08,HCW05,PKN11,CXH09}. Splitting and reshaping of EIT peak has been also studied using two counter-propagating control lasers having same polarization to form standing wave \cite{SMA07,BCP03,AKW02}.
The EIT has also been extensively studied considering multilevel systems and different polarizations of the control and probe lasers in vapor \cite{ABL98,LBL99,ACD02,GWR03,CSB11} as well as in cold atoms \cite{WYJ06,CYC00,EGD01}.
In the presence of magnetic field splitting and reduction of the linewidth of the EIT has been studied as well \cite{WJG05,IFN09}.

The previously studied system \cite{WJG05,IFN09} using $F=1\rightarrow F=2\leftrightarrow F=2$, $\Lambda$-system with single control laser in cold $^{87}$Rb atom including all the magnetic sub-levels in $^{87}$Rb is described based upon the numerical analysis. Here we analytically describe the $\Lambda$-system ($F=1\rightarrow F=2\leftrightarrow F=2$) with two counter-propagating control lasers having orthogonal polarization in $^{87}$Rb vapor. We identify all the sub-systems
and role of individual sub-system and coherence. We also show in what conditions the effect of individual sub-systems and coherence will dominate and how one sub-system influences others.

In order to address this problem we first discuss generalized N-level system of a particular kind as shown in fig. \ref{generalNlevelsystem} and investigate the role of different kind of coherence. We also show the absorption profile of the probe laser with scan of the both, probe and control lasers for three, four, five, and six-level systems. Absorption profile of a probe laser with the scan of the control laser is a way to remove Doppler background. The effect of Doppler averaging is also discussed for these systems.

\section {Theoretical formulation}
\begin{figure}
\begin{center}
\includegraphics[width=7.5cm,height=5.25cm]{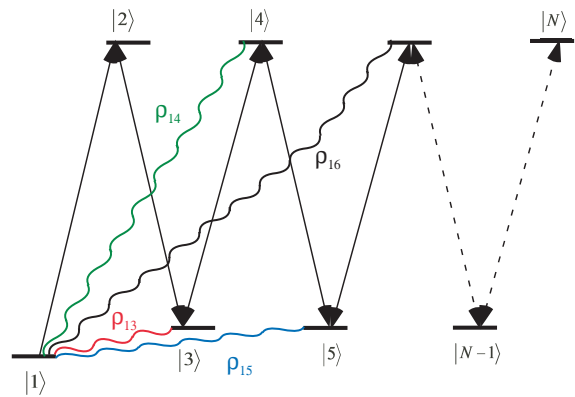}
\caption{(Color online). The energy level diagram for N-level system. The light induced coherence or transfer of coherence (TOC)
between various levels are shown by the curly lines.}
\label{generalNlevelsystem}
\end{center}
\end{figure}

The Hamiltonian of the general N-level system of a type as shown in fig. \ref{generalNlevelsystem} in the rotating frame with rotating wave approximation (R.W.A) is given as,
\begin{eqnarray}
H=\nonumber
\sum_{j=1}^{N-1}\frac{\Omega_{j,j+1}}{2}\Ket{j}\Bra{j+1}+\\
0\Ket{1}\Bra{1}+\sum_{j=2}^{N}\sum_{i=2}^{j}(-1)^{i}\Delta_{i-1,i}\Ket{j}\Bra{j}
\end{eqnarray}
\\
$\Omega_{j,j+1}$ and $\Delta_{j,j+1}$ is Rabi frequency and detuning of the lasers driving levels $\Ket{j}$ $\leftrightarrow$ $\Ket{j+1}$.
In the above Hamiltonian the first summation contains off-diagonal terms, and is about interaction between the adjacent levels with strength of $\Omega_{12}$, $\Omega_{23}$.....$\Omega_{N-1N}$. The second summation contains the diagonal terms which are the energies of various levels in the rotating frame. For example the energy of the $j^{th}$ level is
$\Delta_{12}$-$\Delta_{23}$+$\Delta_{34}$-$\Delta_{45}$....(-1)$^{j}$$\Delta_{j-1j}$. The energy of the ground state, $\Ket{1}$ is taken to be zero.
The probe laser is driving the levels $\Ket{1}$ and $\Ket{2}$ while the control lasers are driving $\Ket{2}\leftrightarrow \Ket{3}\leftrightarrow\Ket{4}....\leftrightarrow \Ket{N}$.
The populations in the various levels and coherence between them is described density matrix, $\rho$.
The diagonal terms of the density matrix describe the population while the off diagonal terms describe the coherences.
The time evolution of density matrix is given by Optical Bloch Equation (OBE) \cite{SCZ97} which is

\begin{equation}
\frac{d\rho}{dt}=i\hbar[\rho,H]
\end{equation}
Time evolution of the population in the various levels is given as,
\begin{align}
\dot{\rho}_{11}&=-\Gamma_1\rho_{11}+\sum^N_{i=2}\Gamma_{i1}\rho_{ii}+\frac{i}{2}\Omega^*_{12}\rho_{12}-\frac{i}{2}\Omega_{12}\rho_{21} \nonumber\\
\dot{\rho}_{22}&=-\Gamma_2\rho_{22}+\sum^N_{i=1,i\neq2}\Gamma_{i2}\rho_{ii}+\frac{i}{2}\Omega_{12}\rho_{21}-\frac{i}{2}\Omega^*_{12}\rho_{12}\nonumber\\
&+\frac{i}{2}\Omega^*_{23}\rho_{23}-\frac{i}{2}\Omega_{23}\rho_{32} \nonumber\\
\vdots\nonumber\\
\dot{\rho}_{jj}&=-\Gamma_j\rho_{jj}+\sum^N_{i=1,i\neq j}\Gamma_{ij}\rho_{ii}+\frac{i}{2}\Omega_{j-1j}\rho_{jj-1}\nonumber\\
&-\frac{i}{2}\Omega^*_{j-1j}\rho_{j-1j}+\frac{i}{2}\Omega^*_{jj+1}\rho_{jj+1}-\frac{i}{2}\Omega_{jj+1}\rho_{j+1j}\nonumber\\
\vdots\nonumber\\
\dot{\rho}_{NN}&=-\Gamma_N\rho_{NN}+\sum^N_{i=1,i\neq N}\Gamma_{iN}\rho_{ii}+\frac{i}{2}\Omega^*_{N-1N}\rho_{N-1N}\nonumber\\
&-\frac{i}{2}\Omega_{N-1N}\rho_{NN-1} \nonumber\\
\end{align}
Where $\Gamma_{ij}$ is the spontaneous decay rate of level $\Ket{i}$ into $\Ket{j}$. For the nonzero value of $\Gamma_{ij}$ the energy of the level $\Ket{i}$ has to be higher than the level $\Ket{j}$. Further the value of $\Gamma_{ij}$ is determined by the dipole matrix element between level $\Ket{i}$ and $\Ket{j}$ . The $\Gamma_{i} \left(=\sum^N_{j=1,j\neq i} \Gamma_{ij}\right)$ is the total decay rate of the level $\Ket{i}$.

The time evolution of the coherence between level $\Ket{1}$ and various other levels is given as,
\begin{align}
\dot{\rho}_{12}&=-\left[\frac{\Gamma_{1}+\Gamma_{2}}{2}-i\Delta_{12}\right]\rho_{12}+\frac{i}{2}\Omega_{12}(\rho_{11}-\rho_{22})\nonumber\\
&+\frac{i}{2}\Omega^*_{23}\rho_{13}\nonumber\\
\dot{\rho}_{13}&=-\left[\frac{\Gamma_{1}+\Gamma_{3}}{2}-i\left(\Delta_{12}-\Delta_{23}\right)\right]\rho_{13}-\frac{i}{2}\Omega_{12}\rho_{23}\nonumber\\
&+\frac{i}{2}\Omega_{23}\rho_{12}+\frac{i}{2}\Omega^*_{34}\rho_{14}\nonumber\\
\vdots\nonumber\\
\dot{\rho}_{1N-1}&=-\left[\frac{\Gamma_{1}+\Gamma_{N-1}}{2}-i\left(\sum_{i=2}^{N-1}(-1)^{i}\Delta_{i-1,i}\right)\right]\rho_{1N-1}\nonumber\\
&-\frac{i}{2}\Omega_{12}\rho_{2N-1}+\nonumber\frac{i}{2}\Omega_{N-2N-1}\rho_{1N-2}+\frac{i}{2}\Omega^*_{N-1N}\rho_{1N}\nonumber\\
\dot{\rho}_{1N}&=-\left[\frac{\Gamma_{1}+\Gamma_{N}}{2}-i\left(\sum_{i=2}^{N}(-1)^{i}\Delta_{i-1,i}\right)\right]\rho_{1N}\nonumber\\
&-\frac{i}{2}\Omega_{12}\rho_{2N}+\frac{i}{2}\Omega_{N-1N}\rho_{1N-1}\nonumber\\
\label{coherence}
\end{align}
In the steady state $\dot{\rho}_{ij}=0$, for all $i$ and $j$. In the case of weak probe limit,
$\rho_{11}\approx1$, $\rho_{22}$, $\rho_{33}$,... $\rho_{NN}$ $\approx0$ and $\Omega_{12}\rho_{23}$, $\Omega_{12}\rho_{24}$,...$\Omega_{12}\rho_{2N}$ $\approx0$.

The set of Eq. \ref{coherence} with the above mentioned approximation gives following,
\begin{equation}
\rho_{1N}\approx\frac{i}{2}\frac{\Omega_{N-1N}}{\gamma_{1N}}\rho_{1N-1}
\label{coh1Np}
\end{equation}
where,
$\gamma_{1j}=\frac{\Gamma_{1}+\Gamma_{j}}{2}-i\displaystyle\sum_{i=1}^{j-1}(-1)^{i+1}\Delta_{i,i+1}$.

\begin{equation}
\rho_{1N-1}\approx\frac{i}{2}\frac{\Omega_{N-2N-1}}{\gamma_{1N-1}}\rho_{1N-2}+\frac{i}{2}\frac{\Omega^*_{N-1N}}{\gamma_{1N-1}}\rho_{1N}
\label{coh1N-1p}
\end{equation}
Eq. \ref{coh1Np} and \ref{coh1N-1p} produce
\begin{equation}
\rho_{1N-1}=\cfrac{\frac{i}{2}\frac{\Omega_{N-2N-1}}{\gamma_{1N-1}}}{1+\frac{1}{4}\frac{\mid\Omega_{N-1N}\mid^2}{\gamma_{1N}\gamma_{1N-1}}}\rho_{1N-2}
\label{coh1N-1}
\end{equation}
Again from Eq. \ref{coherence}
\begin{equation}
\rho_{1N-2}\approx\frac{i}{2}\frac{\Omega_{N-3N-2}}{\gamma_{1N-2}}\rho_{1N-3}+\frac{i}{2}\frac{\Omega^*_{N-2N-1}}{\gamma_{1N-2}}\rho_{1N-1}
\label{coh1N-2p}
\end{equation}
From Eq. \ref{coh1N-1} and \ref{coh1N-2p}
\begin{equation}
\rho_{1N-2}=\cfrac{\frac{i}{2}\frac{\Omega_{N-3N-2}}{\gamma_{1N-2}}}{1+\cfrac{\frac{1}{4}\frac{\mid\Omega_{N-2N-1}\mid^2}{\gamma_{1N-2}\gamma_{1N-1}}}{1+\frac{1}{4}\frac{\mid\Omega_{N-1N}\mid^2}{\gamma_{1N-1}\gamma_{1N}}}}\rho_{1N-3}
\end{equation}
From the above general formula we can write
\begin{eqnarray}
\begin{split}
&\rho_{13}= \\
&{\cfrac{\frac{i}{2}\frac{\Omega_{23}}{\gamma_{13}}}
{1+\cfrac{\frac{1}{4}\frac{\Omega_{34}^2}{\gamma_{13}\gamma_{14}}}
{1+\cfrac{\frac{1}{4}\frac{\Omega_{45}^2}{\gamma_{14}\gamma_{15}}}
{1+\cfrac{\frac{1}{4}\frac{\Omega_{56}^2}{\gamma_{15}\gamma_{16}}}
{1+\cfrac{.}{1+\cfrac{.}{1+\frac{1}{4}\frac{\Omega_{N-1 N}^2}{\gamma_{1 N-1}\gamma_{1 N}}}}}}}}}\\
\end{split}
\label{coh13}
\end{eqnarray}

Again from Eq. \ref{coherence} in steady state
\begin{equation}
\rho_{12}\approx\frac{i}{2}\frac{\Omega_{12}}{\gamma_{12}}+\frac{i}{2}\frac{\Omega^*_{23}}{\gamma_{12}}\rho_{13}
\label{coh12}
\end{equation}
The Eq. \ref{coh13} and \ref{coh12} gives
\begin{eqnarray}
\begin{split}
&\rho_{12}= \\
&\cfrac{\frac{i}{2}\frac{\Omega_{12}}{\gamma_{12}}}
{1+\cfrac{\frac{1}{4}\frac{\Omega_{23}^2}{\gamma_{12}\gamma_{13}}}
{1+\cfrac{\frac{1}{4}\frac{\Omega_{34}^2}{\gamma_{13}\gamma_{14}}}
{1+\cfrac{\frac{1}{4}\frac{\Omega_{45}^2}{\gamma_{14}\gamma_{15}}}
{1+\cfrac{\frac{1}{4}\frac{\Omega_{56}^2}{\gamma_{15}\gamma_{16}}}
{1+\cfrac{.}{1+\cfrac{.}{1+\frac{1}{4}\frac{\Omega_{N-1 N}^2}{\gamma_{1 N-1}\gamma_{1 N}}}}}}}}}
\end{split}
\label{generalformula}
\end{eqnarray}

\subsection {Role of different type of coherence}
\begin{figure}
\begin{center}
\includegraphics[width=7.5cm,height=12cm]{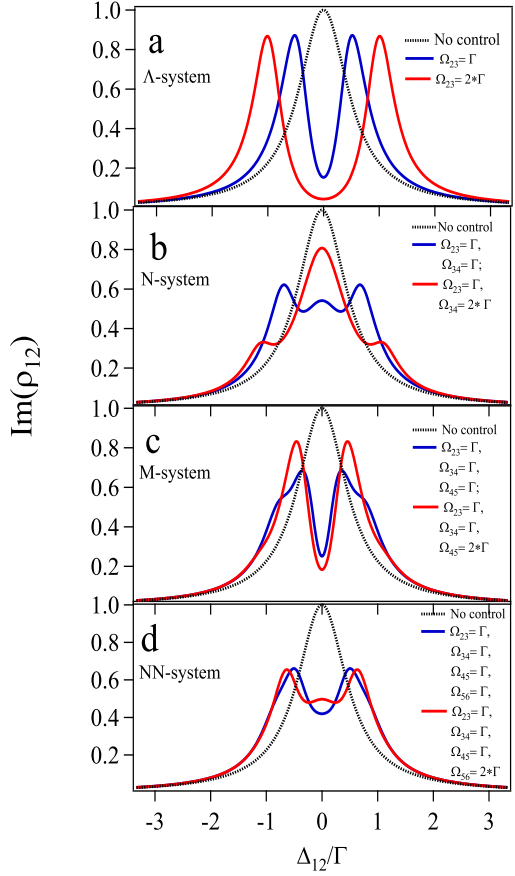}
\caption{(Color online). Absorption profile of probe laser (Im($\rho_{12}$)) with detuning of
same in the presence of control lasers with different combinations of Rabi frequencies.
}
\label{profilenodopplerprobe}
\end{center}
\end{figure}

\begin{figure}
\begin{center}
\includegraphics[width=7.5cm,height=12cm]{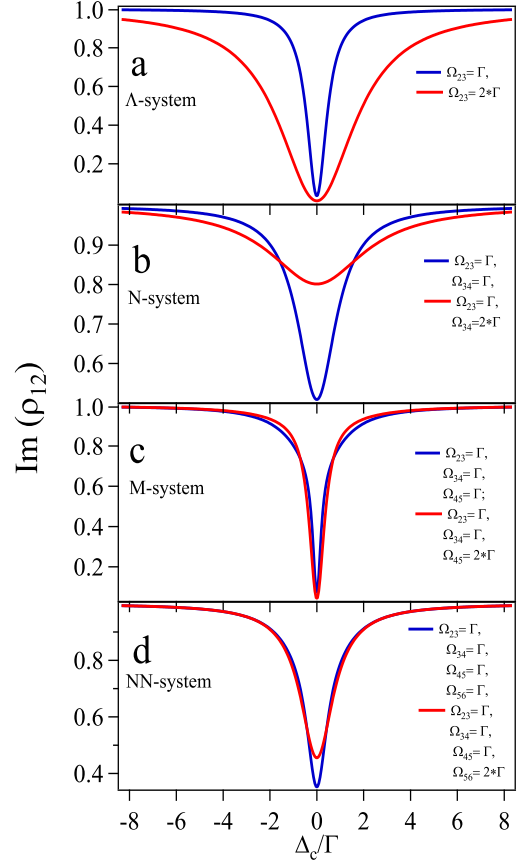}
\caption{(Color online). Absorption profile of probe laser (Im($\rho_{12}$)) with detuning of control lasers in the presence of control lasers with different combinations of Rabi frequency.}
\label{profilenodopplercontrol}
\end{center}
\end{figure}

The above Eq. \ref{generalformula} written in form of series, gives the glimpse of role of the different terms due to TOC. The TOC between level $\Ket{1}$ and any general level $\Ket{i}$ is $\rho_{1i}$ as shown in fig. \ref{generalNlevelsystem}, corresponds to term $\frac{1}{4}\frac{\Omega^2_{i-1i}}{\gamma_{1i-1}\gamma_{1i}}$. The first term $\frac{i}{2}\frac{\Omega_{12}}{\gamma_{12}}$ is the absorption of the probe laser in absence of any control laser. The TOC between level $\Ket{1}$ and $\Ket{3}$ i.e. $\rho_{13}$, corresponds to the term $\frac{1}{4}\frac{\Omega^2_{23}}{\gamma_{12}\gamma_{13}}$, causes reduction of the absorption of the probe laser, which is known as EIT.
The $\rho_{14}$ corresponding to the term $\frac{1}{4}\frac{\Omega^2_{34}}{\gamma_{13}\gamma_{14}}$, causes induced absorption  and known as a EITA. The $\rho_{15}$ corresponding to $\frac{1}{4}\frac{\Omega^2_{45}}{\gamma_{14}\gamma_{15}}$ again reduces the absorption (EITAT) and $\rho_{16}$ corresponding to $\frac{1}{4}\frac{\Omega^2_{56}}{\gamma_{15}\gamma_{16}}$ causes increase absorption (EITATA) and so on. The fig. \ref{profilenodopplerprobe} and \ref{profilenodopplercontrol} shows the absorption profile of a probe laser for three, four, five and six-level system. The three ($\Lambda$-system) and four-level (N-system) are well known but we describe here for the completeness. The fig. \ref{profilenodopplerprobe} and \ref{profilenodopplercontrol} show the absorption profile of the probe laser for the various system with different combinations of Rabi frequencies of the control lasers in the unit of $\Gamma$, and for a decoherence rate between ground states 1/12~$\Gamma$, where $\Gamma$ is decay rate of excited states. The $\Gamma$ in the case of Rb is 2$\pi$$\times$ 6~MHz. The decoherence between the ground state is taken to be 2 $\pi$$\times$ 500~kHz. This decoherence rate includes collisions and the typical linewidth of the lasers.

The fig. \ref{profilenodopplerprobe}a, \ref{profilenodopplercontrol}a which represents $\Lambda$-system, shows that at the line center there is decrease in the absorption of the probe laser in the presence of a control laser (EIT) and with higher Rabi frequency this decreases further. The next fig. \ref{profilenodopplerprobe}b shows that in the presence of another control laser, $\Omega_{34}$ forming N-system, tends to recover the absorption (EITA).
The recovery of absorption is not complete, as at line center, the black dotted line shows more absorption as compared to blue ($\Omega=\Gamma$) and red trace ($\Omega=2 \Gamma$). It is more clear from fig. \ref{profilenodopplercontrol}b, which shows absorption profile of the probe with scan of control lasers. The reason for the incomplete recovery, is higher decoherence rate between level $\Ket{1}$ and $\Ket{4}$ as compared to between level $\Ket{1}$ and $\Ket{3}$, since level $\Ket{4}$ is excited state.
In the presence of one more control laser ($\Omega_{45}$) which forms M-system, the absorption at line center of probe decreases as shown in fig. \ref{profilenodopplerprobe}c, \ref{profilenodopplercontrol}c. In the presence of another control laser ($\Omega_{56}$) which forms NN-system, the absorption tends to recover against transparency created by $\rho_{15}$ but not completely as shown in fig. \ref{profilenodopplerprobe}d \ref{profilenodopplercontrol}d due to higher decoherence rate for $\rho_{16}$ as compared to $\rho_{15}$, since $\Ket{6}$ is excited state.

\subsection{Mutual influence of the two sub-systems}
\begin{figure}
\begin{center}
\includegraphics[width=7.5cm,height=5.25cm]{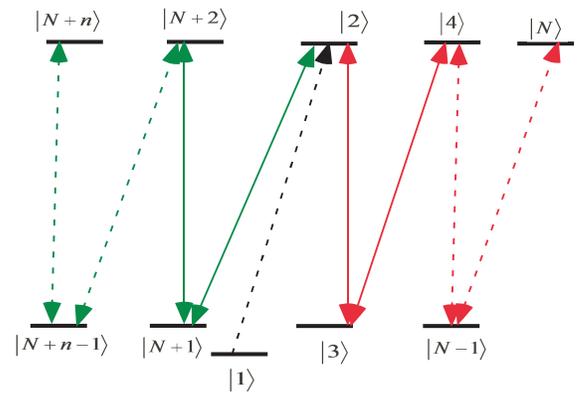}
\caption{(Color online). The energy level diagram of two sub-systems sharing common level $\Ket{2}$. The dotted black arrow is the probe laser. The control lasers shown by red arrows is forming one sub-system while control laser shown with green arrows is forming another.}
\label{influencesubsystem}
\end{center}
\end{figure}

\begin{figure}
\begin{center}
\includegraphics[width=7.5cm,height=6cm]{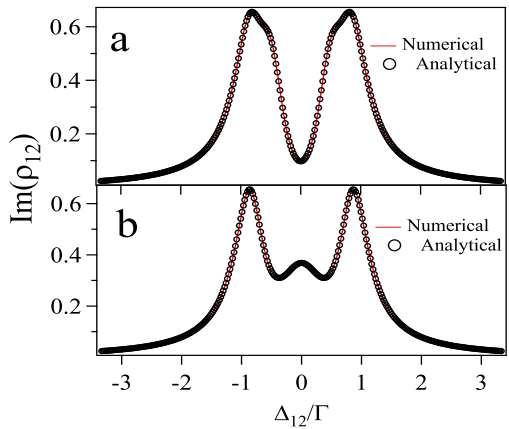}
\caption{(Color online). Comparison between numerical and analytical solution for the absorption profile (Im($\rho_{12}$)) of the probe laser in the presence of control lasers having Rabi frequency $\Gamma$. The decoherence rate between ground states 1/12 $\Gamma$.  (a) Forming one $\Lambda$-system and one M-system. (b) Forming two N-systems}
\label{comnumvsanatwosub}
\end{center}
\end{figure}

In this section we discuss the case where the strong control lasers of two sub-systems share common level as shown in fig .\ref{influencesubsystem}. In this fig the level $\Ket{2}$ is shared by $\Ket{1}$ $\rightarrow$ $\Ket{2}$ $\rightarrow$ $\Ket{3}$$\cdots$$\Ket{N}$ and $\Ket{1}$ $\rightarrow$ $\Ket{2}$ $\rightarrow$ $\Ket{N+1}$$\rightarrow$$\cdots$$\Ket{N+n}$ sub-systems. In the weak probe limit case the absorption of the probe is given by

\begin{align}
&\rho_{12}=\nonumber\\
&\cfrac{\frac{i}{2}\frac{\Omega_{12}}{\gamma_{12}}}{1
+\cfrac{\frac{1}{4}\frac{\Omega^2_{2p}}{\gamma_{12}\gamma_{1p}}}{1+ \cfrac{.}{1+\cfrac{.}{1+\frac{1}{4}\frac{\Omega_{N+n-1N+n}}{\gamma_{1N+n-1}\gamma_{1N+n}}}}}+\cfrac{\frac{1}{4}\frac{\Omega^2_{23}}{\gamma_{12}\gamma_{13}}}{1
+ \cfrac{.}{1+\cfrac{.}{1+\frac{1}{4}\frac{\Omega^2_{N-1N}}{\gamma_{1N-1}\gamma_{1N}}}}}}\nonumber\\
\label{twosubsystemformula}
\end{align}
In the deriving above equation the procedure and approximation is the same as for the Eq. \ref{generalformula}.
In the support of Eq. \ref{twosubsystemformula} we compare this with complete numerical solution for the probe absorption in the presence of control lasers forming a M-subsystem ($\Ket{1}$$\rightarrow$$\Ket{2}$$\rightarrow$$\Ket{3}$$\rightarrow$$\Ket{4}$$\rightarrow$$\Ket{5}$) and a $\Lambda$-subsystem ($\Ket{1}$$\rightarrow$$\Ket{2}$$\rightarrow$$\Ket{N+1}$) in fig. \ref{comnumvsanatwosub}a and two N-subsystems ($\Ket{1}$$\rightarrow$$\Ket{2}$$\rightarrow$$\Ket{3}$$\rightarrow$$\Ket{4}$ and $\Ket{1}$$\rightarrow$$\Ket{2}$$\rightarrow$$\Ket{N+1}$$\rightarrow$$\Ket{N+2}$) in fig. \ref{comnumvsanatwosub}b sharing the level $\Ket{2}$.

\subsection {Effect of the Doppler-broadening}
The effect of Doppler averaging gives interesting modifications of the line-shape of the probe absorption in EIT. \cite{KPW05,BMW08,KPV08}. Linewidth narrowing is an quite interesting fact of it \cite{IKN08,BMW08}. Here we extend this discussion for four and beyond four-level of type N, M and NN-systems. Locking the probe laser and scanning the control lasers all together removes irrelevant Doppler background. Scanning the control lasers all together is quite easy in the case of degenerate energy levels.

For any type of system to see the effect of different TOCs it essential to satisfy two-photon resonance condition ($\Delta_{12}-\Delta_{23}=0$) for a Doppler broadened medium.

\begin{figure}
\begin{center}
\includegraphics[width=7.5cm,height=12cm]{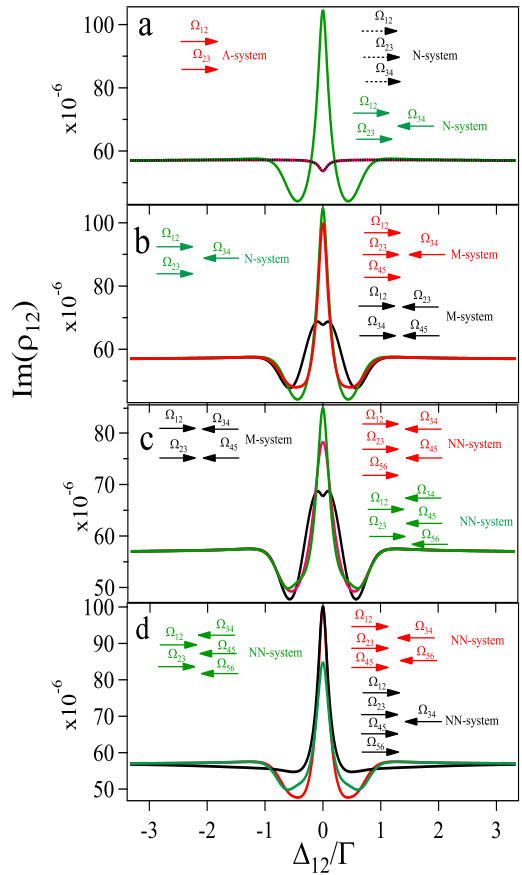}
\caption{(Color online). Absorption profile of probe laser, (Im($\rho_{12}$)) with detuning of probe laser in the presence of control lasers with same Rabi frequencies of $\Gamma$. The propagation direction of individual laser is shown with arrows in the annotation.}
\label{profiledopplerprobe}
\end{center}
\end{figure}

\begin{figure}
\begin{center}
\includegraphics[width=7.5cm,height=12cm]{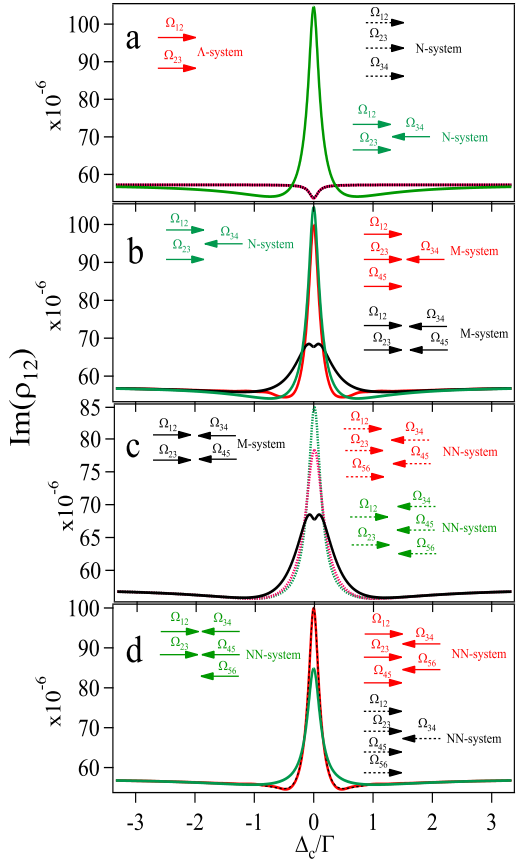}
\caption{(Color online). Absorption profile of probe laser (Im($\rho_{12}$)) with detuning of control laser in the presence of control lasers with same Rabi frequency of $\Gamma$. The co-propagation (Co) and counter-propagation (Cou) configuration of control lasers with respect to propagation of probe laser are added in the annotation.}
\label{profiledopplercontrol}
\end{center}
\end{figure}

First we discuss the effect of Doppler averaging for N-system. For this system we saw in the previous sub-section that in the presence of a control laser, $\Omega_{34}$ the absorption tends to recover but not fully (fig. \ref{profilenodopplerprobe}b, \ref{profilenodopplercontrol}b), while Doppler averaging with counter-propagating $\Omega_{34}$ shows narrow absorption, as shown with green curve in fig. \ref{profiledopplerprobe}a, \ref{profiledopplercontrol}a. The co-propagating, $\Omega_{34}$ control laser only marginally improves the absorption, as shown with black dotted curve in the same figure, but still falls under transparency. The interesting point is the counter-propagating $\Omega_{34}$ control laser gives maximal effect (EITA) of $\rho_{14}$ eventhough the three photon resonance condition ($\Delta_{12}-\Delta_{23}+\Delta_{34}=0$) is not satisfied for moving atoms.

For M-system with co-propagating $\Omega_{23}$, counter-propagating $\Omega_{34}$ and the counter-propagating $\Omega_{45}$ control laser (this satisfies four photon resonance condition, $\Delta_{12}-\Delta_{23}+\Delta_{34}-\Delta_{45}=0$ ) gives maximal effect (EITAT) of M-system as shown with black curve in fig. \ref{profiledopplerprobe}b, \ref{profiledopplercontrol}b. The co-propagating $\Omega_{45}$ give small reduction in the absorption over N-system as shown, with red curve in the same figure. This is the configuration which occurs for the M-system in the doubly driven $\Lambda$-system in $^{87}$Rb described in the next section.

For the NN-system the absorption profile of the probe laser for the various configurations of the control laser propagation is shown in fig \ref{profiledopplerprobe}c, \ref{profiledopplerprobe}d and \ref{profiledopplercontrol}c, \ref{profiledopplercontrol}d. The red curve in fig. \ref{profiledopplerprobe}d, \ref{profiledopplercontrol}d is the configuration which exist in the doubly driven $\Lambda$-system also gives maximal effect of EITATA.

In the case where two sub-systems share a common energy level, the Doppler averaging follows the above discussion.
In order to see the effect of both sub-systems both has to satisfy two-photon resonance condition i.e. $\Delta_{12}-\Delta_{23}=0$ and $\Delta_{12}-\Delta_{2N+1}=0$.
If only one sub-system satisfy the two photon resonance condition the other sub-systems will be just idle.

\section {Various sub-system formed}
In the considered $\Lambda$-system in $^{87}$Rb with the two ground hyperfine states F$_g=1$ and F$_g=2$ and one excited hyperfine state F$_e=2$, the probe laser is $\sigma^-$ polarized and driving F$_g=1$ $\rightarrow$ F$_e=2$ transition. Out of the two control lasers driving F$_g=2$ $\leftrightarrow$ F$_e=2$, one is co-propagating and the other is counter-propagating to the probe laser. This problem is analyzed for the two configurations. One with co-propagating control laser having $\sigma^+$ polarization and counter-propagating control laser having $\sigma^-$ polarization and the other possibility is with the reverse polarization of these two control lasers.

\begin{figure}
\begin{center}
\includegraphics[width=7.5cm,height=15cm]{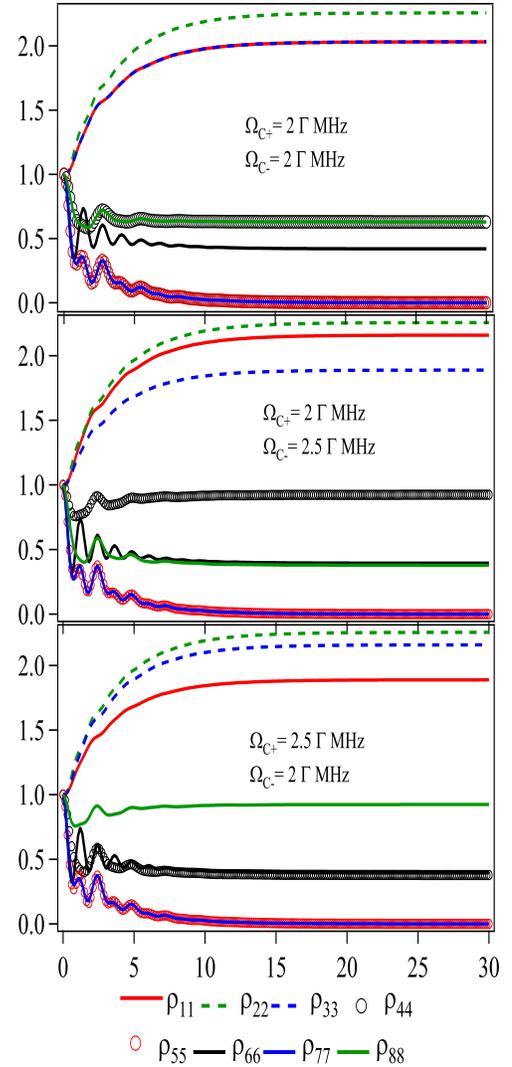}
\caption{(Color online). Population evolution of ground states in the presence of $\sigma^+$ and $\sigma^-$ control lasers.
(a). The Rabi frequencies of $\sigma^+$ and $\sigma^-$ control lasers are 12 and 12 MHz
(b). The Rabi frequencies of $\sigma^+$ and $\sigma^-$ control lasers are 12 and 15 MHz
(c). The Rabi frequencies of $\sigma^+$ and $\sigma^-$ control lasers are 15 and 12 MHz.}
\label{popevo}
\end{center}
\end{figure}

The time evolution of the population of the ground states for different combinations of Rabi frequencies of the two control lasers $\sigma^-$ and $\sigma^+$ is shown in fig. \ref{popevo}. The steady state population of the ground state $\Ket{1}$, $\Ket{2}$ and $\Ket{3}$ is maximum. There are some population left in the states $\Ket{4}$, $\Ket{8}$ and $\Ket{6}$ while the population of states $\Ket{5}$ and $\Ket{7}$ is almost pumped out. Remember the Eq. \ref{generalformula}
was derived with consideration that the population in the other levels are zero which is approximately correct (see fig. \ref{popevo}).

\begin{figure}
\begin{center}
\includegraphics[width=7.5cm,height=5.25cm]{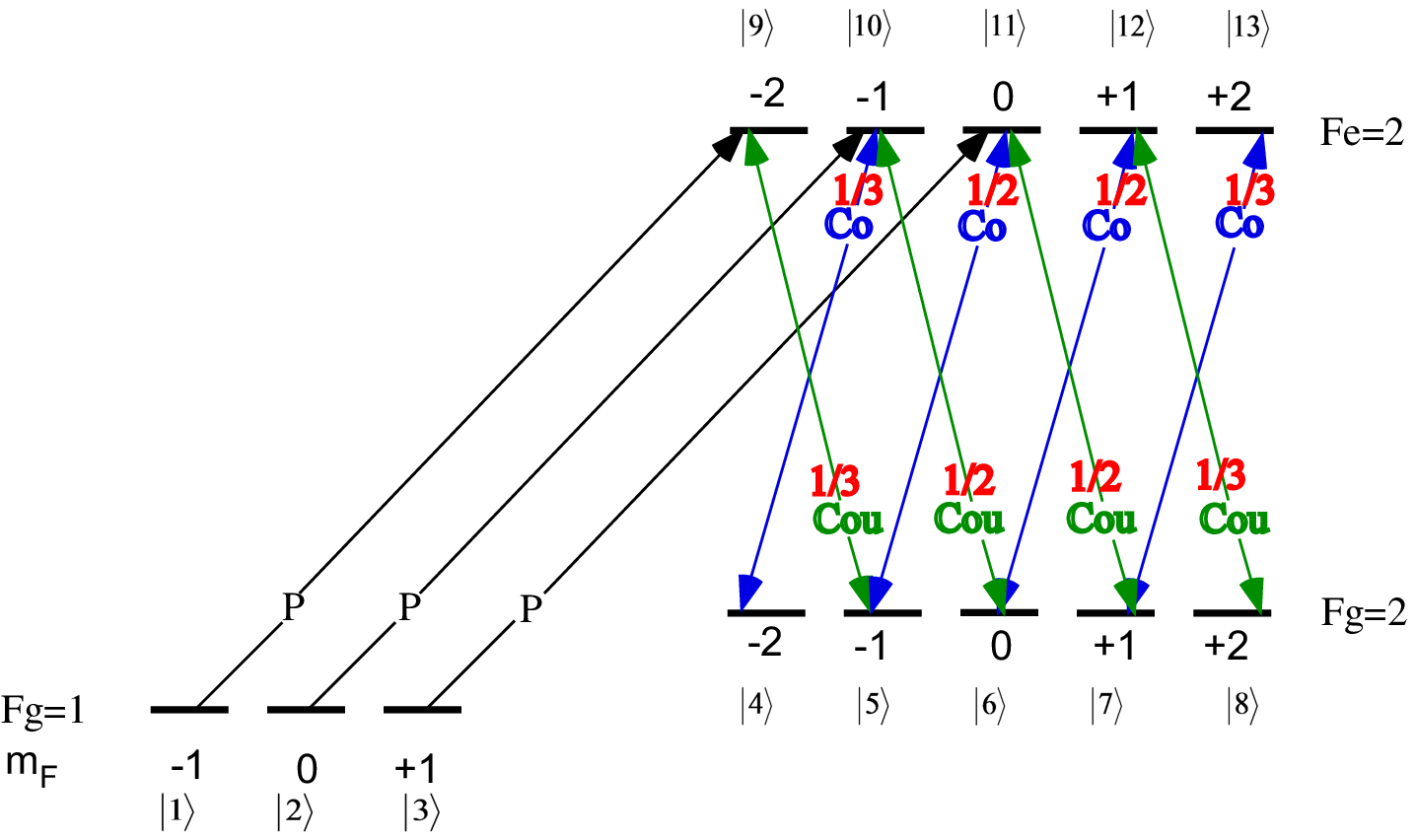}
\caption{(Color online). The energy level diagram for the doubly driven $\Lambda$-system in $^{87}$Rb. Probe (P) laser is $\sigma^-$ polarized,
Co-propagating (Co) control laser is $\sigma^+$ polarized and Counter-propagating (Cou) control laser is $\sigma^-$ polarized.}
\label{levelsigmaplusco}
\end{center}
\end{figure}

\begin{figure}
\begin{center}
\includegraphics[width=7.5cm,height=5.25cm]{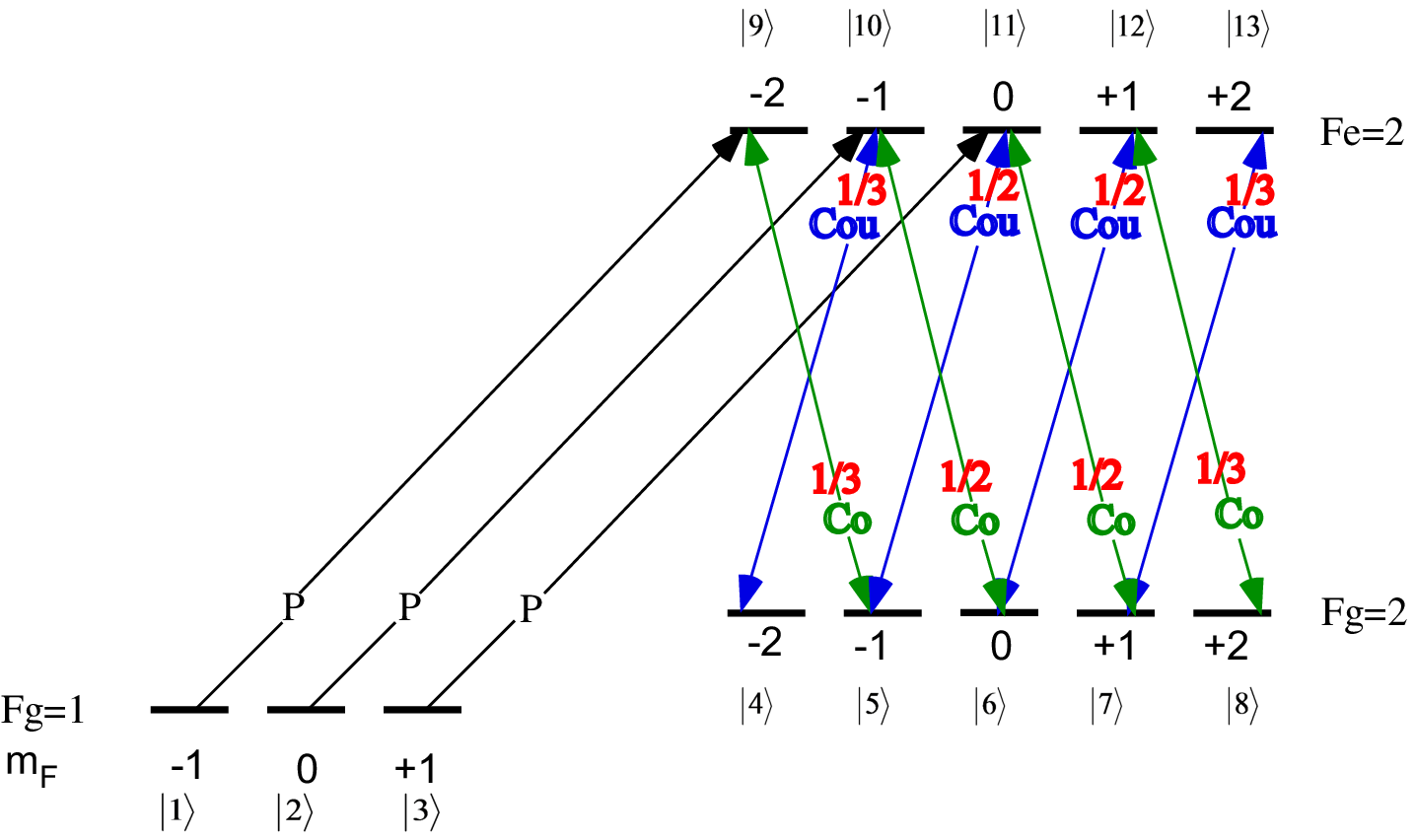}
\caption{(Color online). The energy level diagram for the doubly driven $\Lambda$-system in $^{87}$Rb. Probe (P) laser is $\sigma^-$ polarized,
Co-propagating (Co) control laser is $\sigma^+$ polarized and Counter-propagating (Cou) control laser is $\sigma^-$ polarized.}
\label{levelsigmaminusco}
\end{center}
\end{figure}

The fig. \ref{levelsigmaplusco} and \ref{levelsigmaminusco}  show various sub-systems formed by these two counter-propagating control lasers. The probe laser driving $\Ket{F_g=1; m_F=-1}$ ($\Ket{1}$) $\rightarrow$ $\Ket{F_e=2; m_F=-2}$ ($\Ket{9}$) is forming NN-system i.e. $\Ket{1}$$\rightarrow$ $\Ket{9}$$\rightarrow$$\Ket{5}$$\rightarrow$$\Ket{11}$$\rightarrow$$\Ket{7}$$\rightarrow$$\Ket{13}$. The probe laser driving $\Ket{F_g=1; m_F=0}$ ($\Ket{2}$) $\rightarrow$
$\Ket{F_e=2; m_F=-1}$ ($\Ket{10}$) is forming one $\Lambda$-system ($\Ket{2}$$\rightarrow$ $\Ket{10}$$\rightarrow$$\Ket{4}$) and one M-system ($\Ket{2}$$\rightarrow$ $\Ket{10}$$\rightarrow$$\Ket{6}$$\rightarrow$$\Ket{12}$$\rightarrow$$\Ket{8}$). The M-system and $\Lambda$-system share level $\Ket{10}$.
The probe laser driving $\Ket{F_g=1; m_F=+1}$ ($\Ket{3}$) $\rightarrow$ $\Ket{F_e=2; m_F=0}$ ($\Ket{11}$) forming two N-systems ($\Ket{3}$$\rightarrow$ $\Ket{11}$$\rightarrow$$\Ket{5}$$\rightarrow$$\Ket{9}$ and $\Ket{3}$$\rightarrow$ $\Ket{11}$$\rightarrow$$\Ket{7}$$\rightarrow$$\Ket{13}$). These two N-systems share level $\Ket{11}$.

\begin{figure}
\begin{center}
\includegraphics[width=7.5cm,height=9cm]{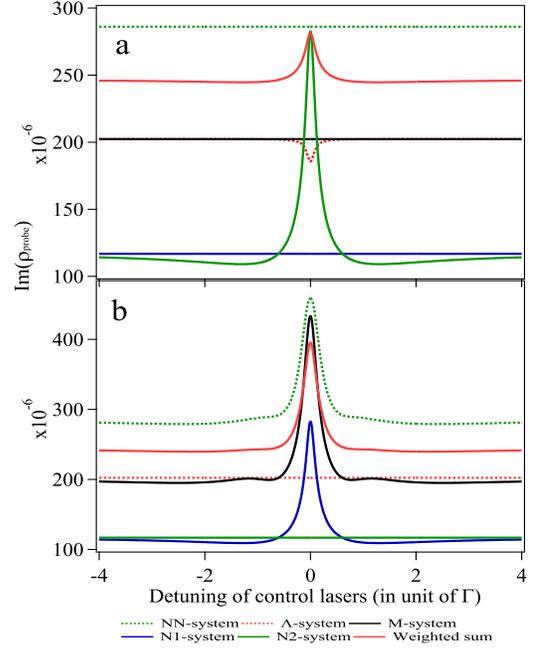}
\caption{(Color online). Absorption profile of probes for the various sub-systems. Probe is $\sigma^-$ polarized (a) $\sigma^+$ control laser is co-propagating while $\sigma^-$ control laser is counter-propagating to the probe laser. (b) $\sigma^-$ control laser is co-propagating while $\sigma^+$ is control laser is counter-propagating. The Rabi frequency of co-propagating control laser is 12~MHz (2~$\Gamma$) and of counter-propagating laser is 15~MHz (2.5~$\Gamma$).}
\label{differentlevels3}
\end{center}
\end{figure}

First we discuss the case of co-propagating $\sigma^+$ and counter-propagating $\sigma^-$control laser as shown in fig. \ref{levelsigmaplusco}. The probe ($\Ket{1} \rightarrow \Ket{9}$) which forms NN-system does not satisfy the two-photon resonance condition and hence will not show effects of any TOCs, as shown with green dotted curve in fig. \ref{differentlevels3}a. The probe ($\Ket{2} \rightarrow \Ket{10}$) forming M-system will also have no effect due to same reason as shown with black curve in the figure, while the same probe forming $\Lambda$-system satisfies the two-photon resonance conditions and shows EIT as shown with red dotted curve. The probe ($\Ket{3} \rightarrow \Ket{11}$) forming two N-systems, only one of them $\Ket{3} \rightarrow \Ket{11}\rightarrow \Ket{5}\rightarrow \Ket{9}$ satisfies two-photon resonance condition and shows EITA as shown with green curve, while the other N-system does not satisfy any two photon resonance condition and will not show EITA as shown with blue curve. The weighted sum with Clebsch-Gordan coefficient and steady state populations of these three probes is shown with red curve.

Now we discuss the case of co-propagating $\sigma^-$ and counter-propagating $\sigma^+$control laser as shown in fig. \ref{levelsigmaminusco}. The probe ($\Ket{1} \rightarrow \Ket{9}$) which forms NN-system satisfies the two-photon resonance condition and hence shows the effect of EITATA, as shown with green dotted curve in fig. \ref{differentlevels3}b. The probe($\Ket{2} \rightarrow \Ket{10}$) forming M-system is having effect of EITAT as shown with black curve, while this probe forming $\Lambda$-system does not satisfy the two-photon resonance condition and shows no EIT as shown with red dotted curve. The probe ($\Ket{3} \rightarrow \Ket{11}$) forming two N-systems, only one of them $\Ket{3} \rightarrow \Ket{11}\rightarrow \Ket{7}\rightarrow \Ket{13}$ satisfies the two-photon resonance condition and shows EITA, as shown with green curve, while the other N-system does not satisfy the two photon resonance condition and will not show EITA as shown with blue curve in the figure. The weighted sum with Clebsch-Gordan coefficient and steady state populations of these three probes is shown with red curve.

\begin{figure}
\begin{center}
\includegraphics[width=7.75cm,height=4.5cm]{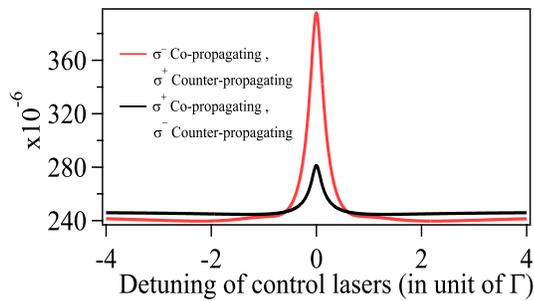}
\caption{(Color online). Comparison of the absorption of a probe for two configurations of control laser polarization. The Rabi frequency of co-propagating control laser is 12~MHz (2~$\Gamma$) and of counter-propagating laser is 15~MHz (2.5~$\Gamma$).}
\label{comparison}
\end{center}
\end{figure}

The comparison of absorption profile for the two configurations, is shown in fig. \ref{comparison}. Clearly with the co-propagating control laser with $\sigma^-$ polarization and counter-propagating control laser with $\sigma^+$ polarization, the absorption of the probe laser is more as compare to the case with reverse polarizations. The linewidth of the absorption profile is sub-natural and smaller than without averaging.

\section{Conclusions}
The general N-level system shows sequence of EIT and EITA due to the various terms in TOCs.
The doubly driven, $F=1\rightarrow F=2\leftrightarrow F=2$, $\Lambda$-system in $^{87}$Rb
can show the effect of many of these TOCs terms depending upon the configurations. The $\sigma^+$ co-propagating and $\sigma^-$ counter-propagating control laser only shows the effect of N-system, means upto $\rho_{14}$, while $\sigma^-$ co-propagating and  $\sigma^+$ counter-propagating shows the effect of all N, M and NN-system, means upto $\rho_{16}$.
The linewidth of the absorption profile of probe with scan of the probe or control laser is subnatural due to Doppler averaging.

\section{Acknowledgment}
We acknowledge Prof. Vasant Natarajan and Sapam Ranjita Chanu for important discussions and providing some
experimental inputs.

\bibliography{eitrefsall}

\end{document}